\documentclass[11pt,preprint2]{aastex}

\slugcomment{Accepted for publication in AJ}

\shorttitle{Overdensities Of EROs In The Fields Of High-$z$ RLQs}
\shortauthors{Wold et al.}

\begin{document}

\title{Overdensities Of Extremely Red Objects In The Fields Of 
High-Redshift Radio-Loud Quasars}

\author{M. Wold\altaffilmark{1}\altaffiltext{1}{JPL Postdoctoral Researcher}\, 
and L. Armus}
\affil{{\em SIRTF} Science Center, 
California Institute of Technology, MS 100-22, Pasadena, CA-91125}
\author{G. Neugebauer} 
\affil{California Institute of Technology, MS 105-24, Pasadena, CA-91125}
\author{T. H. Jarrett}
\affil{{\em IPAC}, California Institute of 
Technology, MS 100-22, Pasadena, CA-91125}
\author{M. D. Lehnert}
\affil{Max Planck Institute for Extraterrestrial Physics, 
Giessenbachstra{\ss}e, 857488, Garching}

\begin{abstract}

We have examined the occurrence of Extremely Red Objects (EROs) in the fields 
of 13 luminous quasars (11 radio-loud and two radio-quiet) 
at $1.8 < z < 3.0$. The average surface density of 
$K_{s}\leq19$ mag EROs is two-three times higher than in large, 
random-field 
surveys, and the excess is significant at the $\approx3\sigma$ level even 
after taking into account that the ERO distribution is highly inhomogeneous. 
This is the first systematic investigation of the surface density of EROs
in the fields of radio-loud quasars above $z\approx2$, and shows that a 
large number of 
the fields contain clumps of EROs, similar to what is seen only in the densest
areas in random-field surveys. The high surface densities and angular 
distribution of EROs suggest that the excess originates in high-redshift galaxy
concentrations, possibly young clusters of galaxies. The fainter EROs at 
$K_{s}\ga 19$ mag show some evidence of being more clustered in the immediate 
20 arcsec region around the quasars, suggesting an association with the 
quasars. Comparing with predictions from spectral synthesis models, we find 
that if the $K_{s}\approx19$ mag ERO excess is associated with the quasars at 
$z\approx2$, their magnitudes are typical of $\gtrsim L^{*}$ passively 
evolving galaxies formed at $z\approx3.5$ ($\Omega_{m}=0.3$, 
$\Omega_{\Lambda}=0.7$, and $H_{0}=70$ km\,s$^{-1}$\,Mpc$^{-1}$). Another 
interpretation of our results 
is that the excess originates in concentrations of galaxies at $z\approx1$ 
lying along the line of sight to the quasars. If this is the case, the EROs 
may be tracing massive structures responsible for a magnification bias of the 
quasars.
\end{abstract}


\section{Introduction}

The age and space density of the oldest and most massive galaxies at each 
epoch provide the strongest constraints on models of galaxy formation. This is
one of main reasons that there has been so much focus on Extremely Red Objects
(EROs). It is the high-redshift systems that are
most important for testing models, and the ERO selection criterion is
designed to find old, passively evolving galaxies at $z \gtrsim 1$.
Because the 4000 {\AA} break in evolved galaxies 
shifts from optical to near-infrared at 
$z \gtrsim 1$, EROs are defined in terms of optical to 
near-infrared color, 
with two of the most used definitions being $R-K\geq5.0$ and $I-K\geq4.0$
mag. 

The co-moving volume density, or even the number density, of massive, evolved
galaxies at $z\gtrsim 1$ can place strong constraints on models of galaxy 
formation.
Whereas monolithic collapse 
models \citep{eggen62,larson74,tg76,lp03} predict pure 
luminosity evolution of early-type galaxies from the period of formation 
until the present-day, hierarchical formation models \citep{wr78,bcf96,
kauffmann96,baugh98,spf01} predict significant density evolution 
from $z\approx1$ to $z\approx0$. Testing the models has proven to 
be difficult, and the results are not yet conclusive. Some surveys
have found a deficit of early-type galaxies at $z\approx1$ 
\citep{barger99,smith02,roche02} whereas others have not 
\citep{daddi00b,cimatti02,im02}. Part of the disagreement may be caused by
the strong clustering of EROs on the sky \citep{daddi00,firth02,roche02},
and large fields must be surveyed in order to overcome this problem. 

The ERO definition also selects dusty starburst galaxies 
whose spectral 
energy distribution can mimic those of old, passively evolving galaxies. 
It is still unclear how large a fraction of the ERO 
population consists of dusty starburst galaxies. Several methods are being
used to separate the two galaxy types, including 
photometric \citep{mannucci02,smail02,miyazaki02}, spectroscopic 
\citep{cimatti02,saracco03}, and those based on morphology 
\citep{moriondo00,st01,yt03}. The results so far seem to suggest that roughly 
two-thirds of the EROs at $K\leq19$ mag are old, passively evolving galaxies.
However, it is also clear that the situation is more complex 
than just a simple distinction into old, evolved galaxies and dusty
starbursts since systems with old populations and residual star formation
also appear in ERO samples \citep{yt03}.  

In this paper, we are concerned with EROs in the fields of high-redshift 
quasars. 
Some early studies 
serendipitously found EROs in the fields of $z>2$ radio galaxies and quasars
\citep{mccarthy92,hr94,dey95}. Although some of these EROs were later shown 
to lie in the foreground of the radio galaxies \citep{gd96}, probing the 
fields around high-redshift radio-loud AGN (active galactic nuclei) has been a 
promising method of finding high-redshift galaxies. Luminous
quasars and radio galaxies are often associated with high-density galaxy 
environments \citep{eyg91,hl91,hg98,wold00,best00,venemans02,best03}, and the 
rich environments are often interpreted as clusters or groups of galaxies 
hosting the AGN. By combining a targeted search with a color criterion like 
that of EROs, high-redshift early-type galaxies can be effectively filtered 
out from the numerous foreground galaxies. This becomes very important at
high redshifts, since even for a rich cluster at 
$z\approx1$, most of the galaxies observed using a single broad-band filter 
are foreground or background galaxies \citep{dickinson97}.
A few surveys have found that EROs seem to be more common in high-redshift 
AGN fields than in the general field 
\citep{chapman00,cimatti00,hall01,sgs02},
and there are individual AGN fields which show striking overdensities of EROs
\citep{thompson00,liu00,haines01}.

Even though galaxy excess in the fields of AGN are often being
interpreted as groups or clusters physically associated with the AGN 
(spectroscopically confirmed in a few cases 
\citep{dickinson97,deltorn97}), gravitational lensing is also known to cause 
correlations (and anti-correlations) between 
high-redshift quasars and optically bright 
foreground galaxies \citep{webster88,hlf90,benitez95,benitez97,ni00,myers03}. 
It is possible that a similar situation could arise between high-redshift 
quasars ($z\gtrsim 2$) and EROs, because the EROs are likely to be tracers of 
massive structures which may boost the fluxes of distant background quasars
by gravitational lensing.

The aim with our present study is to test whether EROs are more common in 
the fields of $z\gtrsim 2$ quasars. To our knowledge, this is the first 
systematic survey to probe the surface density of EROs in the fields 
of quasars at $z>2$, and preliminary results were reported by 
\citet{wold03}. The quasars in our sample are selected from 
the catalogs of \citet{barthel88} and \citet{hb93}, and form a subset of a 
larger sample of 40 designed to study host galaxies (Armus et al.\ 
in preparation). The 40 sample quasars are nearly evenly divided between 
radio-loud and radio-quiet systems with comparable $V$ magnitudes and 
redshifts. The subsample of 13 analyzed here were selected randomly from
the sample of 40 and followed up with wide-field near-infrared
and optical imaging. The sample is listed in Table~\ref{table:t1}.
In this table, we have also listed the radio loudness, defined in terms
of the ratio of the flux densities at rest-frame 5 GHz and 2500 {\AA},
$S_{5 GHz}/S_{2500}$ \citep{sw80}. According to the definition by 
\citet{kellermann89}, a quasar is radio-loud if $S_{5 GHz}/S_{2500} > 10$,
and thus two of the 13 quasars are classified as radio-quiet (one of these
might be radio-intermediate instead of radio-quiet) whereas the rest are 
radio-loud. In calculating the radio loudness, we used the $K_{s}$-band fluxes
of the quasars as found from our data, and the 
radio fluxes at 5 GHz listed in the catalog of \citet{barthel88} 
unless otherwise is noted in the table. The $K_{s}$-band fluxes were 
converted to rest-frame 2500 {\AA} by assuming a power-law spectrum,  
$S_{\nu}\propto \nu^{-\alpha}$, where $\alpha$ was found from the median
quasar energy distribution by \citet{elvis94}. The conversion to rest-frame
5 GHz radio flux was done by assuming the same power-law form of the spectrum,
but using spectral indices listed in the catalog by \citet{barthel88}, unless
otherwise is noted in the table.

\begin{deluxetable}{lllrr}
\tablecaption{The quasar sample. \label{table:t1}}
\tablehead{
\colhead{Quasar} & 
\colhead{$\frac{S_{5 GHz}}{S_{2500}}$} &  
\colhead{$z$} & 
\colhead{$R$} & 
\colhead{$K_{s}$} \\
\colhead{(1)} & \colhead{(2)} & \colhead{(3)} & \colhead{(4)} & \colhead{(5)}
}
\startdata
0741$+$169 		     & 3880\tablenotemark{a} &  1.894  &  7500 & 
3280 \\
0758$+$120 		     & 1690 &  2.660  &  3600 & 2360 \\
0830$+$115 		     & 2050 &  2.974  &  4550 & 2720 \\
0927$+$217		     & 130  &  1.830  &  3600 & 3280 \\
0941$+$261 		     & 5750 &  2.910  &  2400 & 2240 \\
1056$+$015  & $<$14\tablenotemark{b}&  2.650  & 11100 & 3260 \\ 
1354$+$258 		     & 1080 &  2.032  &  2500 & 6800 \\
1456$+$092 		     & 2590 &  1.991  &  3900 & 3200 \\
1629$+$680 		     & 4070 &  2.475  &  6100 & 3780 \\
1630$+$374  & $<$6\tablenotemark{b} &  2.037  &  3600 & 2920 \\ 
1702$+$298 		     & 18130&  1.927  &  4000 & 3340 \\
2150$+$053 		     & 3330 &  1.979  &  2000 & 3040 \\
2338$+$042 		     & 4915 &  2.594  &  3600 & 2880 \\
\enddata
\tablecomments{
(1) IAU designation. 
(2) Radio loudness defined as the ratio of the flux densities at rest-frame
5 GHz and 2500 {\AA} \citep{sw80}. (3) Redshift. (4)--(5) Total 
exposure time in seconds in the $R$- and $K_{s}$-filters.}
\tablenotetext{a}{Radio flux density and spectral index 
from the Green Bank radio survey by
\citet{becker91}.}
\tablenotetext{b}{Upper limit based on a 3$\sigma$ detection 
in the NRAO-VLA Sky Survey \citep{condon98} and an assumed spectral index 
of $\alpha=0.7$.}
\end{deluxetable}

The structure of the paper is as follows. In the next section, we
describe the observations and data reduction.
Section~\ref{section:sec3} deals with the construction of object catalogs and
addresses their incompleteness. 
In Section~\ref{section:sec4} we present the results of the analysis.
The surface density of EROs is calculated and the ERO excess
evaluated, carefully taking into account their strong clustering. We
also investigate the $K_{s}$-band number counts and the 
radial distribution of EROs around the quasars. 
In Section~\ref{section:sec5}, we discuss two different interpretations of the 
results, and the conclusions are drawn in Section~\ref{section:sec6}. 

All magnitudes are given in the Vega system, and our assumed cosmology has 
$\Omega_{m}=0.3$, $\Omega_{\Lambda}=0.7$, and $H_{0}=70$ 
km\,s$^{-1}$\,Mpc$^{-1}$. 

\section{Observations and data reduction}

The data consist of wide-field $R$- and $K_{s}$-band
images taken with the Palomar 200-inch telescope using the prime-focus 
instruments COSMIC\footnote{Carnegie Observatories Spectrograph and Multiple 
Imaging Camera} \citep{cosmic} and 
PFIRCAM\footnote{Prime Focus Infrared Camera} \citep{jarrett94a}.
The COSMIC instrument is equipped with a 2048$\times$2048 
Tektronix CCD with a pixel scale of 0.2856 arcsec, and the PFIRCAM employs 
a 256$\times$256 HgCdTe array with a pixel scale of 0.494 arcsec. The 
resulting fields of view in $R$ and $K_{s}$ are therefore 
9.7$\times$9.7 and 2.1$\times$2.1 arcmin$^{2}$, respectively.

Images were obtained during several observing runs from September 1995 to
July 2001. We used a 9- or 27-point dither pattern for the $K_{s}$ images, and
on-chip integration times of 4 sec. The number of co-adds was set so that the 
total integration time at each dither point was 20--30 sec, and the offsets 
between individual frames were typically 10--15 arcsec. The $R$-band 
exposures were divided into integrations of 300--600 sec each, with an offset 
of 10--20 arcsec between each exposure. The total integration time in $R$ and 
$K_{s}$ for each field is listed in Table~\ref{table:t1}. 

Flat-field images in $K_{s}$ were made by taking the median of 
dark-subtracted quasar images. Typically, a flat-field was made for each
dither sequence, and this was found to be sufficient in most cases. After 
flat-fielding, the images were sky subtracted, corrected for bad pixels and 
registered using the {\sc dimsum} package \citep{sed95} within 
{\sc iraf.}\footnote{{\sc iraf} is distributed by the National Optical 
Astronomy Observatories, which are operated by the Association of 
Universities for Research in Astronomy, Inc., under cooperative agreement 
with the National Science Foundation.} 

The sky subtraction in {\sc dimsum} is performed by subtracting a running sky
frame made by taking the 
median of nine neighboring images. After the sky subtraction, masks are created
for objects and cosmic rays and the masks are 
utilized to make a first pass combined image. The first pass image is 
then used to make a new and improved mask by including fainter 
objects. The new object mask is used in the 
final pass where the sky subtraction is repeated and the images aligned and 
combined. At this final step, we block replicated the images 
by a factor of two so that sub-pixel shifts were used for the aligning.
The resulting images therefore have a pixel scale of 0.247 arcsec. 

The $R$-band images were reduced in {\sc iraf} in a standard manner. The bias
level was subtracted either by using a bias frame or by using a 3rd order 
Legendre to fit the overscan region which was then subtracted from each row. 
A flat-field was formed for each quasar field by taking the median of the 
quasar images, or of the quasar images in combination with exposures of other
fields taken near in time. Typically 15--30 frames were used for this,
and all emission $\approx2\sigma$ above the average background was masked
out.
This method rejected cosmic rays, stars and galaxies visible in 
the quasar images and yielded a high quality flat-field frame. The 
flat-fielded images were then aligned using 20--40 stars 
in each frame, and averaged using {\it imcombine} with a high 
threshold set to remove bad pixels that had been flagged by giving them high,
discrepant values. The combined $R$ images were aligned with, and put to the 
pixel scale of, the $K_{s}$ images using the {\sc iraf} tasks {\em geomap} 
and {\em geotran}. 

The flux calibration of the $R$-band images was accomplished using standard
stars from \citet{landolt92}, and the $K_{s}$-data were calibrated using 
standard stars from \citet{persson98}. The conditions were photometric, and 
the transparency good throughout most of the runs. The final photometric 
status was checked by the amount of variation within the individual exposures 
of the same source and by following standard stars throughout the night. 
For fields that were imaged during non-photometric conditions, calibration 
images were obtained later during photometric conditions
to allow for an accurate flux calibration. During the 
photometric nights, the zeropoint varied by less than 0.1 mag. 
For a subset of the images, we checked our absolute photometry against
the Two Micron All Sky Survey (2MASS) point source catalog 
(Skrutskie et al.\ 1997; 
Cutri et al.\footnote{http://www.ipac.caltech.edu/2mass/releases/second/
doc/explsup.html}). 
A very good agreement, typically $<0.1$ mag difference, was found between our 
$K_{s}$-magnitudes and the 2MASS $K_{s}$-magnitudes for bright stars 
of $K_{s} \lesssim 15$ mag. 

The full width at half maximum (FWHM) of the seeing in the $K_{s}$-images 
is 0.8--1.0 arcsec, and in 
the $R$-images, typically 1.0--1.5 arcsec. The raw number 
counts of galaxies with detection significance $\geq3\sigma$ 
turns over at $K_{s}\approx20$ mag, although this varies 
somewhat from field to field due to the different exposure times. 
In the $R$-band the corresponding turn-over in the number counts is 
$R\approx24.5$ mag. The 3$\sigma$ detection limits in the images are 
$K_{s}\approx21.5$ and $R\approx26$ mag. We address the completeness of the 
images in the next section. 

\section{Construction of object catalogs}
\label{section:sec3}

\subsection{Source detection}

For object detection and photometry we used the {\sc sextractor} software 
package \citep{ba96}. In order to have the same FWHM of the seeing
in both $R$ and $K_{s}$ we convolved the $K_{s}$-band images with a Gaussian
function using the {\sc iraf} task {\it gauss} prior to object detection.
For each field, object catalogs were constructed based on the detections in 
the $K_{s}$-band image. During the detection process, we used a Gaussian 
convolution filter with a FWHM matched to the seeing and 
a detection threshold of 1.5$\sigma$. Since the images were block
replicated, we set the minimum number of pixels above 
the threshold to 20 in order to avoid spurious detections and
cosmic rays over an area of less than 5 pixels in the original images. 
The result of the detection process was inspected visually in order to 
ensure that
no obvious objects were missed, and that no false detections were entered into
the catalogs. Saturated objects and objects lying close to the 
image boundaries were rejected from the catalogs.

\begin{figure}
\includegraphics{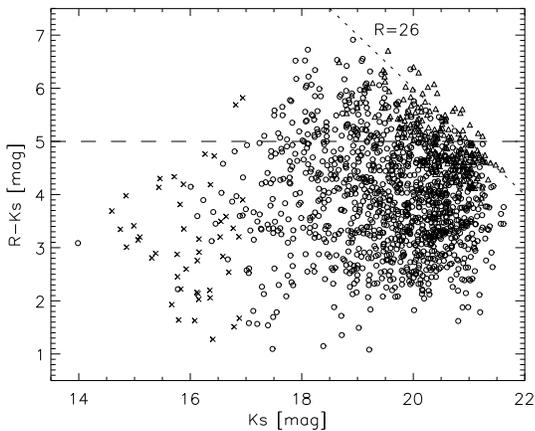}
\caption{The circles show objects detected at a 
significance $\geq 5\sigma$ in the $K_{s}$-filter. The triangles are 
objects with a detection significance $<3\sigma$ in the $R$-filter where a 
lower limit to the 
$R-K_{s}$ color has been calculated assuming a 3$\sigma$ detection in $R$. 
The crosses are objects classified as stars. The horizontal line
indicates our ERO definition, and the diagonal line the $R=26$ mag limit.
The scatter of objects above the diagonal line arises from the varying
depths of the $R$-band images. \label{figure:fig1}}
\end{figure}

Because of the dithering technique used in the $K_{s}$-band,
the exposure time falls off toward the edges in the final combined
images.
Since the PFIRCAM field of view is relatively small, we included lower 
exposure regions, but rejected the regions where the exposure time is less 
than 25 \% of the total, i.e.\ where the signal-to-noise drops by more 
than 50 \%. This causes the 
rms background noise to increase toward the edges, resulting in a number
of spurious detections. In order to avoid such spurious detections, we 
used the exposure maps produced by {\sc dimsum} as weights 
in {\sc sextractor}. 
This allows for an adjustment of the detection threshold according to the 
noise level.

After the $K_{s}$-catalogs were made, matched $R$-band catalogs were 
created by running {\sc sextractor} in double image mode. This produces a 
catalog of fluxes measured in the $R$-band image, but using the $K_{s}$-band 
image for detection. For total magnitudes, we used the {\sc sextractor} 
{\it best} magnitudes which are evaluated using adaptive apertures, or using 
isophotal corrections if a neighbor is suspected of biasing the adaptive 
aperture magnitude \citep{ba96}. For determining colors, we used aperture
magnitudes, and chose an aperture diameter of two times the FWHM of the 
seeing, corresponding to typically 9--10 pixels, or 2.2--2.5 arcsec.

For each $K_{s}$-selected object, we calculated the detection 
significance using the uncertainty in the flux as measured by {\sc sextractor}
which takes into account 
the Poissonian nature of the counts and the standard deviation
in the background counts. In addition to this, we added in quadrature the 
error arising from the fluctuating background level across the image
\citep{best00}. 
In order for a $K_{s}$-selected object to be included in the catalog we 
required the 
detection significance within the aperture to be $\geq 5\sigma$. 
If the flux measured within 
the corresponding aperture in the $R$-band was significant at the 
$\geq 3\sigma$ level, we evaluated the $R-K_{s}$ color, and if $<3\sigma$, a
lower limit to the $R-K_{s}$ color was calculated assuming a $3\sigma$ 
detection in $R$. 

In order to separate stars and galaxies, we used the star-galaxy classifier in
{\sc sextractor} which assigns a probability between 0 and 1 to an object
according to how likely it is to be a star. The star-galaxy classifier works
well, but becomes less reliable at faint magnitudes. We therefore performed
star-galaxy separation only at $K_{s}\leq17$ mag, rejecting objects with
a star-galaxy class of $\geq 0.85$ from the catalogs. The star-galaxy 
separation was based on the detections in the $K_{s}$ images. 

\subsection{Completeness}

To be able to address the completeness of the $K_{s}$-band catalogs, we 
added artificial galaxies to the images using the {\sc iraf} task
{\em mkobjects}. For a given input magnitude, we added galaxies with 
both deVaucouleurs and exponential disk profiles, in total 
$\approx 1500$ galaxies distributed over the 13 fields. The scale-lengths
of the artificial galaxies were randomly selected from the FWHM of
the objects in the catalogs, and the images convolved with the same
Gaussian function as was used to put the $K_{s}$-band images to the 
seeing of the $R$-band images. After this, the images were 
processed through {\sc sextractor} using the same detection criteria 
as described above.

The simulations show that for input objects of $K_{s}=18$ mag,
96 \% of the galaxies are recovered, where the missing 4 \% is due
to blending with other objects. At $K_{s}=19$ mag the completeness drops
to 90 \%, and at $K_{s}=20$ mag the catalogs are $\approx40$ \% complete.
The numbers were evaluated assuming a 50:50 mix of deVaucouleurs and 
exponential disk profiles.

\section{Results}
\label{section:sec4}

\subsection{The surface density of EROs}

We select EROs in the quasar fields using the criterion $R-K_{s} \geq 5.0$
mag. This includes old, passively evolving galaxies at $z \geq 0.9$, 
but also dusty starbursting galaxies. With only two filters, and at the 
resolution of our images, it is not possible to distinguish between these two 
galaxy types. There will also be some contamination by faint, red stars 
in our ERO sample as discussed below.
We have chosen the $R-K_{s}\geq5.0$ mag 
selection criterion because of two reasons.
Firstly, because selecting galaxies with red colors enables us to 
eliminate a large part of the numerous foreground galaxies at $z<1$ while at 
the same time producing a big enough sample to perform meaningful statistics 
on. Secondly, because several random-field surveys employing the 
$R-K_{s}\geq5.0$ mag criterion exist with which we can compare the ERO 
surface density. Our areal coverage in $R$ and $K_{s}$ is limited by the 
small field of view of the PFIRCAM instrument, so the field 
counts cannot be estimated from the quasar images themselves. 

The color-magnitude diagram of all the $K_{s}$-selected objects 
is shown in Fig.~\ref{figure:fig1}. Above our $R-K_{s} \geq 5.0$ mag 
selection there are 253 objects, and 105 of these have a detection significance
in $R$ of three or less, i.e.\ the lower limit on their $R-K_{s}$ color 
is $\geq5.0$ mag. The {\sc sextractor} star-galaxy classifier finds that two 
of 
the EROs at $K_{s}\leq17$ mag are stars, and these are excluded from the ERO 
sample, resulting in a total of 251 galaxies
(including the $\leq3\sigma$ detections in $R$ having a lower limit of
$R-K_{s}\geq5.0$ mag). 

\begin{figure*}
\includegraphics{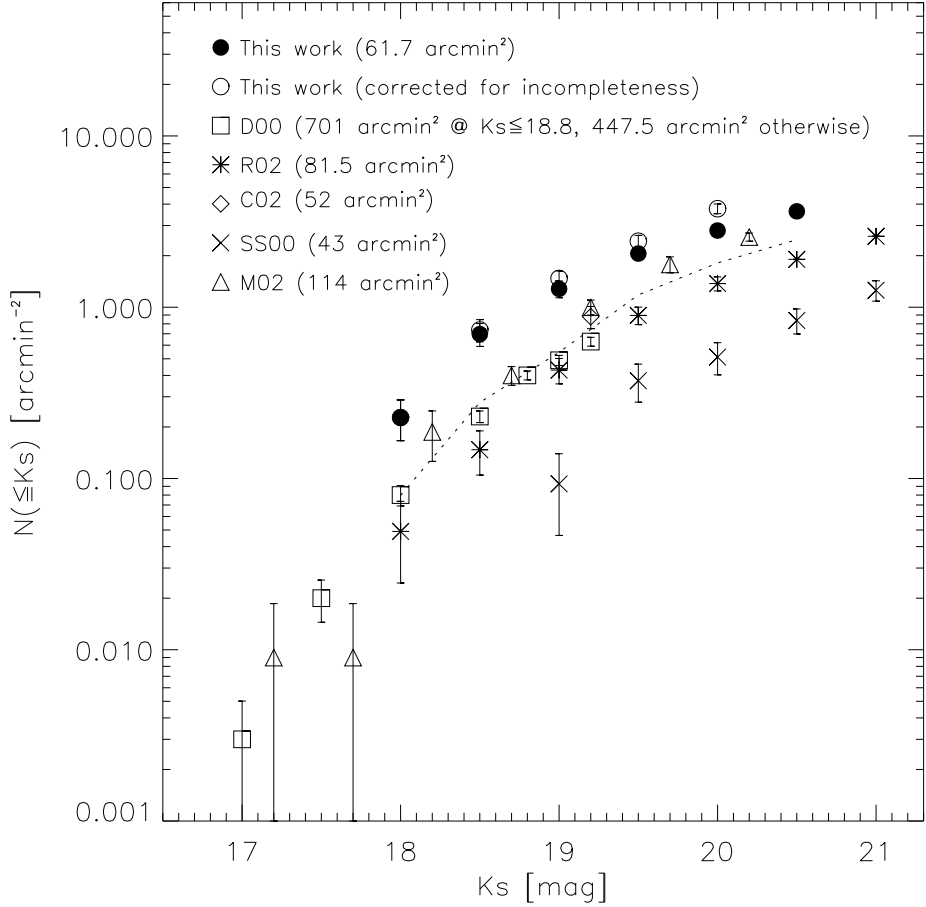}
\caption{The cumulative surface density of EROs in the 
quasar fields compared
with literature surveys in $R-K$ or $R-K_{s}$ covering wide, random fields 
(Roche et al.\ 2002 (R02); Cimatti et al.\ 2002 (C02); Miyazaki et al.\ 2002
(M02); Scodeggio \& Silva 2000 (SS00); Daddi et al.\ 2000a (D00)).
The average of the literature
values (excluding the SS00 counts) is indicated by the dotted line. Error bars
are Poissonian. \label{figure:fig2}}
\end{figure*}

In Fig.~\ref{figure:fig2}, we have plotted the average cumulative surface 
density of EROs in the quasar fields together with ERO 
surface densities found in
the random-field surveys of \citet{daddi00}, \citet{ss00}, \citet{cimatti02}, 
\citet{roche02}, and \citet{miyazaki02}. Note that we have not compared with
field counts derived from other colors (e.g.\ $I-K$ or $I-H$) since
they may bias toward later-type
galaxies with prolonged star formation \citep{mccarthy01,yt03}.
Note also that Miyazaki et al.\ define EROs as having
$R-K_{s}(AB) \geq 3.35$ mag, which, according to Miyazaki et al.,
corresponds to $R-K_{s}(Vega) \geq 4.95$ mag. 

The literature surveys cover wide, random 
areas and should therefore be representative of the ERO counts in the general
field. The field counts 
agree well with each other except for the counts in the
Chandra Deep Field South by \citet{ss00}, which are consistently 
lower than the other. This is most likely caused by the 
inhomogeneity of the ERO distribution coupled with the relatively small 
(and contiguous) area surveyed by Scodeggio \& Silva. 
Fig.~\ref{figure:fig2} demonstrates that the ERO counts in 
our quasar fields are 
systematically higher than those in the field. At $K_{s}\leq19$ mag, where our 
counts are complete, the average surface density of EROs is two--three 
times larger than the average of the field values (excluding those in 
the Chandra Deep Field South). 
According to the average of the random-field surveys, as indicated by
the dotted line in Fig.~\ref{figure:fig2}, we expect 
$34\pm14$ EROs at $K_{s}\leq 19$ mag over an area equal to the 
total area of our survey. (The error in this number takes into account the 
strong clustering of EROs as explained below.)
The observed number of EROs in the quasar fields at $K\leq19$ mag
is 79, hence there is a significant excess above the field counts.
At $K_{s} \ga 20$ mag we still see a clear excess in the quasar fields, 
but our counts become strongly affected by incompleteness.
 
\begin{figure}
\includegraphics{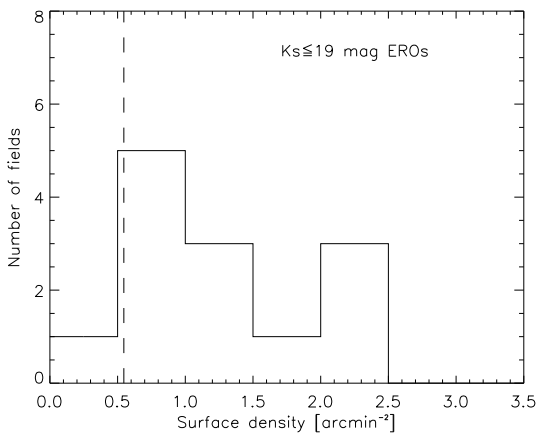}
\caption{Histogram of raw ERO surface densities at 
$K_{s}\leq19$ mag in the quasar fields. The vertical, dashed line indicates 
the average of the random-field surveys, $n=0.546$ arcmin$^{-2}$.
\label{figure:fig3}}
\end{figure}

The number of EROs per quasar field and the average cumulative surface density
as a function of $K_{s}$-magnitude is listed in Table~\ref{table:t2}. Some 
fields are seen to be very rich in EROs, such as 1456$+$092 and 2150$+$053, 
with surface densities four--five times larger than the field values. These
fields have ERO excesses in every magnitude bin. 
Other fields have surface densities consistent with the field counts, like 
0741$+$169 and 1629$+$680. The majority of the fields, however, have surface 
densities in between these two extremes. This is illustrated in 
Fig.~\ref{figure:fig3}, where a histogram of the ERO surface densities at 
$K_{s}\leq19$ mag is shown. If the quasar fields were random with respect to 
the ERO distribution, we would expect a normal distribution centered on 
$\approx 0.546$ arcmin$^{-2}$, which is the average surface density in the 
general field. This is not seen, instead there is a
tail toward high surface densities, with as many as four fields 
(30 per cent of the sample) having a surface density larger than 
three times the field value, and seven fields where the surface density is
twice the field value. 
The combination of the very rich and the moderately rich fields gives an 
overall surface density well above the average for the random-field
surveys. 

\begin{deluxetable}{llccccccc}
\tablecaption{The cumulative number of EROs in the quasar fields as a function
of total $K_{s}$ magnitude. \label{table:t2}} 
\tablehead{
\colhead{Quasar}  & \colhead{Area}  & \colhead{$\leq 18.0$} & 
\colhead{$\leq 18.5$} & \colhead{$\leq 19.0$} & \colhead{$\leq 19.5$} & 
\colhead{$\leq 20.0$} & \colhead{$\leq 20.5$}\\
\colhead{\phantom{Quasar}} & \colhead{arcmin$^{2}$} & \colhead{mag} & 
\colhead{mag} & \colhead{mag} & \colhead{mag} & \colhead{mag} & \colhead{mag}
}
\startdata
0741$+$169 & 5.22 &  0  & 1  &  4 &  7  &  9  & 10 \\
0758$+$120 & 5.23 &  2  & 5  &  6 &  8  & 12  & 16 \\
0830$+$115 & 5.28 &  1  & 3  &  7 & 10  & 14  & 18 \\
0927$+$217 & 4.19 &  1  & 4  &  8 & 12  & 18  & 22 \\
0941$+$261 & 4.16 &  0  & 2  &  5 &  8  & 14  & 17\\
1056$+$015 & 5.13 &  0  & 2  &  4 &  6  &  8  & 16 \\
1354$+$258 & 5.07 &  0  & 3  &  5 & 10  & 14  & 17 \\
1456$+$092 & 4.34 &  2  & 4  & 10 & 12  & 15  & 19 \\
1629$+$680 & 5.09 &  2  & 3  &  4 & 11  & 13  & 21 \\
1630$+$374 & 4.30 &  1  & 2  &  2 &  5  &  5  & 5 \\
1702$+$298 & 5.13 &  1  & 1  &  4 &  9  & 14  & 15 \\
2150$+$053 & 4.21 &  2  & 7  & 10 & 13  & 16  & 21 \\
2338$+$042 & 4.36 &  2  & 6  & 10 & 16  & 21  & 27 \\
 \\
$N_{qso}$  & 61.7 & 14  & 43 & 79 & 127 & 173 & 224  \\ 
$n_{qso}$ (arcmin$^{-2}$) & \nodata & 
0.23$\pm$0.06 & 
0.70$\pm$0.11 & 
1.28$\pm$0.14 & 
2.06$\pm$0.18 & 
2.80$\pm$0.21 &
3.63$\pm$0.25 \\
$n_{stars}$ (arcmin$^{-2}$) & \nodata & 
0.11$\pm$0.04 & 0.17$\pm$0.05 & 0.25$\pm$0.06 & 0.44$\pm$0.08 & 
0.47$\pm$0.09 & 0.59$\pm$0.10 \\
\enddata
\tablecomments{The numbers in this table are based on raw counts.
The bottom three lines show the total number of EROs in the 13 fields, 
$N_{qso}$, the resulting average surface density, $n_{qso}$, with its 
associated Poissonian error, and the expected surface density of 
stars, $n_{stars}$, with its associated Poissonian error.
According to random-field 
surveys, the expected number of $K_{s}\leq 19$ mag EROs over 5 arcmin$^{2}$, 
the typical size of one quasar field, is two--three.}
\end{deluxetable}

\subsection{Contamination by red stars}

Of the 13 quasars fields, 10 lie at high 
galactic latitudes $b>35^{\circ}$, so the contamination by stars is likely
to be small. Using a color-cut of $R-K_{s}\geq5.0$ mag will eliminate most
field stars because of their bluer color, but L- and M-type dwarfs will 
be red enough to make it into our ERO sample.  
Based on $R-K'$ and $J-K'$ colors, \citet{mannucci02} find that 
9 \% of their $K' \leq 20$ mag ERO sample (at $b=-23^{\circ}$) is likely to 
consist of stars. 

In order to estimate the contamination by stars, we used the stellar 
distribution model developed by \citet{jarrett92} (see also 
\citet{jarrett94b}). The model is an extension of the \citet{bs80} optical 
star count model to the near-infrared, and its performance with respect to 
2MASS near-infrared star counts is demonstrated by \citet{cambresy02}. For 
every quasar position, we used the model to predict the expected number of 
stars having $R-K_{s}\geq5.0$ mag. The predicted average surface density of 
stars for the whole ERO sample is listed on the bottom line of 
Table~\ref{table:t2}. These numbers show that 40--50 \% of the brighter EROs 
at $K_{s}\leq18$ mag might be stars, but that at fainter magnitudes the 
contamination decreases to 15--20 \%. Note that this is an overestimate 
because the ERO counts suffer from incompleteness at fainter magnitudes. The 9
\% contamination estimated by \citet{mannucci02} based on two colors therefore
seems to agree well with the model predictions. We thus expect roughly 10 \% 
of our EROs to be red stars. Since we cannot use the small quasar fields 
for estimating the field counts of EROs, we compare the counts in the quasar
fields with number densities of EROs from random-field surveys found in the 
literature. Since the random-field surveys are also contaminated by
stars, a contamination of 10--20 \% does not alter our conclusions.

The three lowest galactic latitude fields in our sample are
0741$+$169, 0758$+$120 and 0830$+$115 at $b=20-27^{\circ}$. For these
fields the model predicts 2.9, 2.6, and 1.6 stars at $K_{s}\leq19$ mag, 
respectively. Note also that these are not among the richest fields in our 
sample.

\subsection{Significance of the ERO excess} 

In this section we evaluate the significance of the ERO excess,
taking into account the large variance in the ERO field counts caused by 
their strong clustering. 
The significance of the excess is given by
\begin{equation}
\sigma_{exc} = \frac{(N_{qso}-N_{field})}{\sigma}
\label{equation:eq1}
\end{equation}
\noindent
\citep{ylc99}, where $N_{qso}$ is the total number of EROs in the quasar 
fields brighter than a given $K_{s}$-magnitude and 
$N_{field}$ is the number of EROs in the random field scaled to the total 
area of the quasar fields, 61.7 arcmin$^{2}$. The error in the field counts 
expected over an area of 61.7 arcmin$^{2}$ is denoted $\sigma$. Ideally 
$N_{field}$ should have been estimated from the edges of the quasar fields, 
but our fields are too small for this. Instead we use the average of the 
surface densities found by \citet{daddi00}, \citet{roche02}, and 
\citet{miyazaki02}. 

Because ERO positions are correlated on the sky, the variance is given by 
\begin{equation}
\sigma^{2} = N_{field} \left(1+N_{field}A_{\omega}C\right)
\label{equation:eq2}
\end{equation}
\noindent
\citep{daddi00}, where $A_{\omega}$ is the amplitude of the ERO angular 
two-point correlation function. The integral constraint which arises from
the fact that the survey area is limited is given by $A_{\omega}C$, and 
Daddi et al.\ find that $C$ can be approximated by $58 \times Area^{-0.4}$. 
We use this approximation and find that $C=11.15$ for the 13 quasar fields. 
The approximation is valid
for a contiguous area, but serves our purpose well even if the 13 fields are 
non-contiguous. In any case, $C=11.15$ is likely to be an overestimate since 
our fields are widely separated and not correlated in any way. We 
use the amplitudes, $A_{\omega}$, found by Daddi et al.\ and Roche et al.\ 
for $K_{s} \leq 19$ mag and by Roche et al.\ for $K_{s}\leq19.5$ mag, and 
calculate $\sigma$. 

The result of the calculations for the entire quasar sample on average
is shown in Table~\ref{table:t3}, where it can be seen that the excess at 
$K_{s}\leq18$, 18.5 and 19 mag is significant at the $\approx3\sigma$ level. 
Because of incompleteness, the excess is not significant at fainter 
$K_{s}$-levels, but if a correction is made for the incompleteness, the excess
appears to be significant also at fainter $K_{s}$ magnitudes, as shown in the 
last column of Table~\ref{table:t3}. 

\begin{deluxetable}{lrrrrr}
\tablecaption{Significance of the average ERO excess in the quasar fields. 
\label{table:t3}}
\tablehead{
\colhead{$K_{s}$} & \colhead{$N_{qso}$} & \colhead{$N_{field} \pm \sigma$} & 
\colhead{$N_{exc}$} & \colhead{$\sigma_{exc}$} & \colhead{$\sigma_{exc}^{corr}$
} \\
\colhead{(1)} & \colhead{(2)} & \colhead{(3)} & \colhead{(4)} & \colhead{(5)} 
& \colhead{(6)}
}
\startdata
$\leq 18.0$   & 14  &   4.9$\pm$3.4       & 9.1  & 2.7 & \nodata \\
$\leq 18.5$   & 43  &  17.2$\pm$9.5       & 25.8 & 2.7 & 3.0 \\
$\leq 19.0$   & 79  &  33.7$\pm$14.3      & 45.3 & 3.2 & 4.0 \\
$\leq 19.5$   & 127 &  72.8$\pm$31.1      & 54.2 & 1.8 & 2.5 \\
$\leq 20.0$   & 173 & 111.9$\pm$34.4      & 61.1 & 1.8 & 3.5 \\
$\leq 20.5$   & 224 & 153.1$\pm$46.4      & 70.9 & 1.5 & \nodata \\
\enddata
\tablecomments{(1) $K_{s}$ magnitude limit. (2) Total number of 
EROs in the 13 quasar fields. (3) The expected total number 
in the 13 fields based on the average of the literature counts, where $\sigma$
is the uncertainty taking into account that the
ERO distribution is inhomogeneous. (4) Excess above the field counts.
(5)--(6) Significance of the ERO excess, calculated using 
Eq.~\ref{equation:eq2}. All numbers, except those in column (6), were 
evaluated using raw counts.}
\end{deluxetable}

\subsection{Radial distribution}

\begin{figure}
\includegraphics{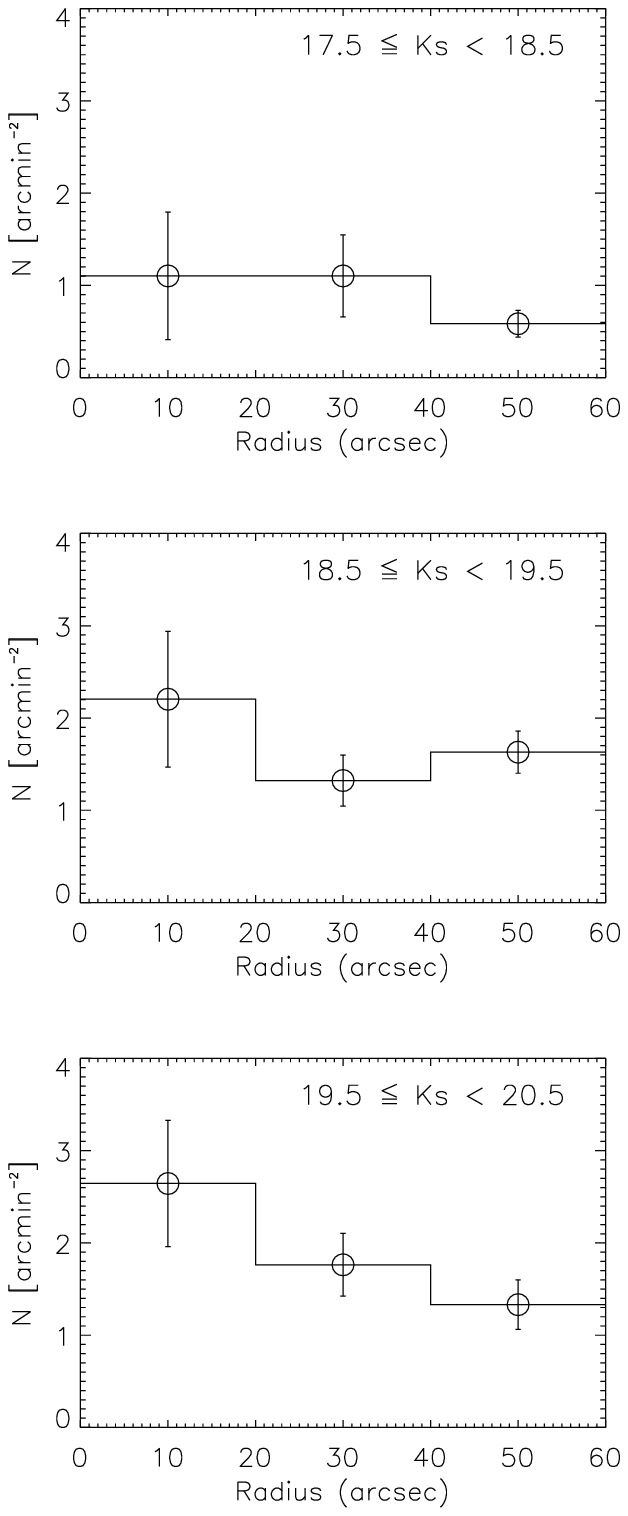}
\caption{Mean radial distribution of EROs in the 13 quasar 
fields based on raw counts. The inner bin is centered on the quasars. 
The error bars show the standard deviation in the mean. \label{figure:fig4}}
\end{figure}

If the EROs are associated with the quasars either through a surrounding
cluster or a magnification bias we might expect to see a trend with
projected quasar separation.
In order to examine the radial distribution of EROs, we 
evaluated the average surface density in three annuli 
centered on the quasars, for three different magnitude intervals.
The result is shown in Fig.~\ref{figure:fig4}. The 
brighter EROs at $17.5 \leq K_{s} < 19.5$ mag are seen to be uniformly
distributed over the whole field, but the fainter ones at 
$K_{s} \ga 19.5$ mag seem to cluster more in the 
immediate 20 arcsec region around the quasars. 
Note that this is not caused by the mosaicing technique 
in the $K_{s}$ filter. The exposure times across the inner two annuli are 
uniform, with the
third annulus being on the average only slightly less well exposed than the 
other two. Using completeness simulations we find that the variation
in the completeness level across the three annuli is at most a few per cent,
i.e. consistent with what we expect from object crowding and blending.
Therefore, there appears to be a real trend toward central concentration
of the fainter EROs. \citet{cimatti00} looked for central concentrations
of $K_{s}$-selected galaxies toward $z\approx1.5$ radio-loud AGN. They
did not find any significant trends, but noted that the regions closest to the
AGNs had a higher number density of galaxies than in the general field. 

\subsection{The surface density of $R-K_{s}\geq6.0$ mag EROs}

For completeness, we also evaluated the surface density of $R-K_{s}\geq6.0$
mag EROs. Using the redder color criterion gives a much
smaller sample, 14 EROs at $K_{s}\leq19$ mag and 31 EROs at $K_{s}\leq20$
mag. The resulting surface densities are thus $0.23\pm0.06$ and 
$0.50\pm0.09$ arcmin$^{-2}$ at $K_{s}\leq19$ and $\leq20$ mag, respectively
(Poissonian errors). The surface density of EROs at $K_{s}\leq19$ mag is 
therefore 3--5 times higher than that found in the field surveys by 
\citet{thompson99} and \citet{daddi00} ($0.039\pm0.016$ and $0.07\pm0.01$ 
arcmin$^{-2}$, respectively). We also note that the 
surface density of $R-K_{s}\geq6.0$ EROs in our quasar fields agrees with the 
ERO surface densities found in radio-loud AGN fields at $z=1-1.5$ by 
\citet{hg98} and \citet{cimatti00}. 

\subsection{$K_{s}$-band counts}

\begin{figure*}
\includegraphics{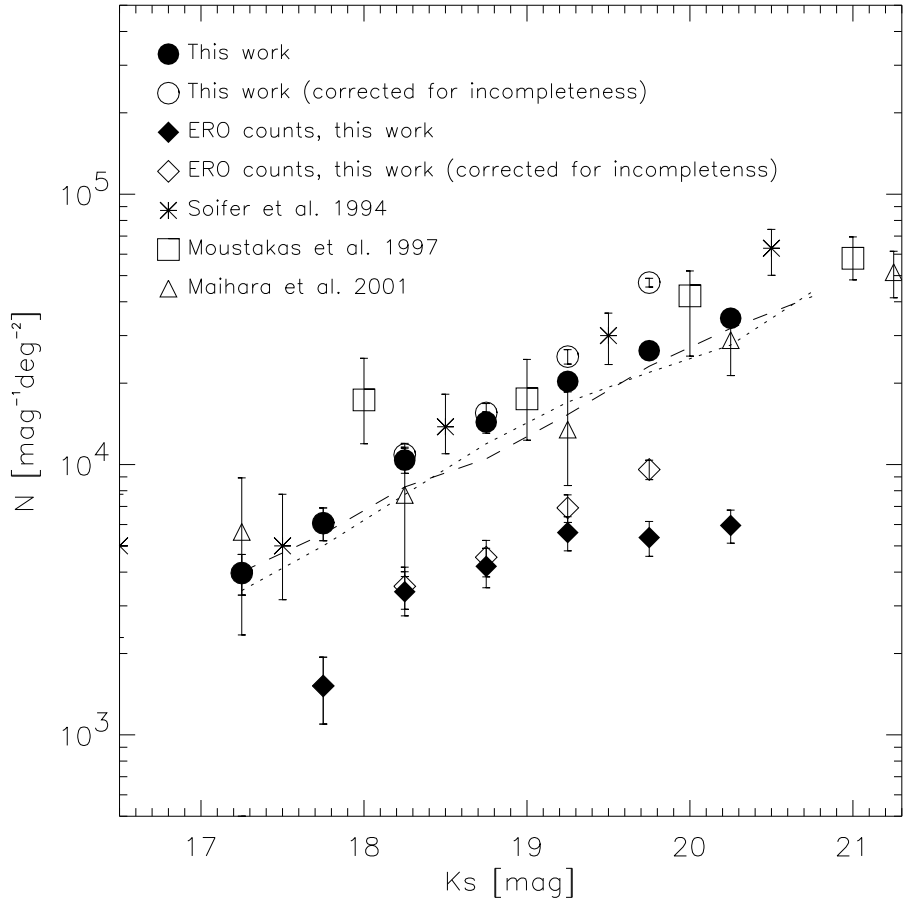}
\caption{Number counts of $K_{s}$-selected galaxies 
(circles) and EROs (diamonds) detected at a significance greater than 5.
We have compared with a selection of deep $K$-band surveys from 
the literature as shown by the other symbols. The dotted and dashed
lines show the literature averages compiled by \citet{hall98} and 
\citet{best03}, respectively. \label{figure:fig5}}
\end{figure*}

Whereas there is a clear excess of EROs in the quasar fields, 
the overall $K_{s}$-counts show only a slight excess at fainter 
magnitudes, $K_{s}\gtrsim 18.5$ mag.  
This is seen in Fig.~\ref{figure:fig5} where we have plotted the average 
$K_{s}$-band number counts together with counts from the surveys by 
\citet{soifer94}, \citet{moustakas97}, and \citet{maihara01}, as well as 
the literature averages compiled by \citet{hall98} and 
\citet{best03}. There are large field-to-field variations
in the random-field $K_{s}$-band counts, and even though the counts in 
the quasar fields
appear to lie above the literature averages at fainter $K_{s}$ magnitudes,
they are still consistent with the counts from the surveys by e.g.\ 
\citet{soifer94}, \citet{moustakas97}, and \citet{maihara01}. 
We also find that the $R$-band number counts are 
consistent with the counts in the general field \citep{metcalfe95,hogg97}. 
Overplotted in Fig.~\ref{figure:fig5}
are also the number counts of EROs in our fields, demonstrating,
as expected, that non-ERO galaxies clearly dominate 
at all magnitudes. 

Table~\ref{table:t4} shows the fraction of EROs with respect to the whole 
$K_{s}$-selected sample of galaxies. For comparison, we have also listed
the ERO fractions given by \citet{daddi00}, \citet{ss00}, and \citet{roche02}, 
the latter calculated from their table~2. 
In the quasar fields, the EROs make up typically 20--25 
\% of all the $K_{s}$-selected galaxies, whereas the field surveys 
find $\lesssim 10$ \%. We note that this behavior is what we expect if the 
EROs are tracing the early-type population in high-redshift clusters of 
galaxies since the fraction of early-type galaxies in rich clusters is known 
to be different from the field \citep{dressler80}.
This also illustrates the richness of EROs in 
the quasar fields, and suggests that the EROs are efficiently tracing
overdensities not readily apparent from the $K_{s}$ number counts alone. 
Clearly, the color-selection is very effective in eliminating foreground
contamination. 

\begin{deluxetable}{llcccc}
\tablecaption{ERO fractions among $K_{s}$-selected galaxies. \label{table:t4}}
\tablehead{
\colhead{$K_{s}$} & \colhead{This work} & \colhead{D00} & \colhead{SS00} & 
\colhead{R02} \\
\colhead{(1)} & \colhead{(2)}& \colhead{(3)}& \colhead{(4)}& \colhead{(5)}  
}
\startdata
$\leq 18.0$  & 0.14$\pm$0.04 & 0.05    & \nodata & \nodata \\
$\leq 18.5$  & 0.22$\pm$0.04 & 0.08    & \nodata & \nodata \\
$\leq 19.0$  & 0.25$\pm$0.03 & 0.12    & 0.02$\pm$0.01 & \nodata \\
$\leq 19.5$  & 0.26$\pm$0.03 & \nodata & 0.06$\pm$0.02 & 0.12$\pm$0.02 \\
$\leq 20.0$  & 0.24$\pm$0.02 & \nodata & 0.06$\pm$0.01 & 0.14$\pm$0.01 \\
$\leq 20.5$  & 0.22$\pm$0.02 & \nodata & \nodata       & \nodata \\
\enddata
\tablecomments{(1) Magnitude limit. (2) Fraction of EROs evaluated using raw
counts. (3)--(5) ERO fractions found by Daddi 
et al.\ 2000a (D00), Scodeggio \& Silva 2000 (SS00), and Roche et al.\ 2002 
(R02). The uncertainties are Poissonian.}
\end{deluxetable}

\section{Discussion}
\label{section:sec5}

We have demonstrated above that the quasar fields have a significant excess of
EROs above what is expected from the general field population of EROs, taking
into account their strong clustering. It appears that the quasar fields contain
clusters or dense concentrations of red galaxies, and their radial 
distribution suggests that the fainter EROs may be related to the quasars. 
However, another possibility is that the EROs are related to structures 
in the foreground of the quasars. We discuss each of these possibilities
below. Note that our sample is dominated by radio-loud quasars, and that
we restrict the following discussion to radio-loud quasars only.

\subsection{`Clusters' of EROs associated with the quasars}

An argument in favor of the ERO excess being at the quasar redshifts is 
that luminous radio-loud quasars and radio galaxies at $z\lesssim 1$ are 
known to reside 
in regions of enhanced galaxy density, interpreted as clusters or groups of 
galaxies physically associated with the AGN \citep{hg98,best00}. Since 
powerful AGN with supermassive black holes may trace the densest regions, 
it is natural to assume that this is true also at $z\gtrsim2$. For example, 
\citet{cimatti00} find excess EROs in radio-loud AGN fields 
at $z\approx1.5$ and \citet{best03} find that red galaxies 
cluster around powerful radio sources at $z\approx1.5$. Our study pushes this
out to $z\gtrsim2$. 

If the ERO excess originates in dense environments associated with the 
quasars, the majority of the EROs must lie at $z\approx2$ since this is the 
redshift of most of the quasars in our sample. 
It is therefore appropriate to discuss whether the color and the 
brightness of the EROs are consistent with being evolved galaxies at 
$z\approx2$. We have used the {\sc p{\'e}gase} models of \citet{frv97} to 
examine both the apparent magnitude and the $R-K_{s}$ color of passively 
evolving galaxies at $z\approx2$ (assuming no extinction by dust, and an 
initial metallicity of 1 \% solar value). The apparent $K_{s}$ magnitude of 
$L^{*}$ galaxies at $z=2$ was found by calculating the scale 
factor between the observed flux of the model and that of a present-day 
$L^{*}$ galaxy. The present-day $L^{*}$ galaxy was assumed to have 
$M_{B}^{*}=-20.3$ mag \citep{loveday92} and an age of 14 Gyr. 

The standard model in which the stars form in a 1 Gyr long burst at $z>4$ 
has an $R-K_{s}$ color at $z=2$ which is too red for most of our EROs. In 
order for this model to produce a color of $R-K_{s}=5-6$ mag at $z=2$, the 
formation redshift has to be $z\approx3.5$. The $K_{s}^{*}$ magnitude of
such a galaxy observed at $z=2$ is 19.2 mag. Since the star formation in 
this model stops abruptly after 1 Gyr, this model represents the galaxy
which reddens most quickly. Therefore, $K_{s}=19.2$ mag is the brightest 
magnitude we can expect for an $L^{*}$ galaxy at $z=2$. 
If the $K_{s}\approx19$ mag ERO excess is associated with passively evolving 
galaxies at $z\approx2$, they must therefore be $\gtrsim L^{*}$ galaxies. 
Other models which have $R-K_{s}=5-6$ mag at $z=2$ include those with 
exponentially decreasing star formation rates.
For instance, a galaxy formed at $z=6$ with an exponentially decreasing star 
formation rate of $\tau=0.5$ Gyr has $R-K_{s}=5-6$ mag 
at $z=2$, and $K_{s}^{*}=19.6-19.7$ mag. If the EROs are described by
this latter model, they must have small amounts of residual
star formation in them. 

The recent spectroscopic work of \citet{vandokkum03} has shown that red 
galaxies at $z>2$ can be very bright. Five of their six galaxies were 
found to lie at $2.4\leq z \leq3.5$ and to have $K_{s}$-magnitudes in 
the range 19.2--19.9 mag. These galaxies were selected by the $J-K_{s}>2.3$ 
mag criterion \citep{franx03}, whereas our galaxies are selected on the basis 
of having $R-K_{s}\geq5.0$ mag. Both criteria are designed to find 
galaxies with evolved 4000 {\AA} breaks, but at different 
redshifts. However, these two selection techniques will not 
necessarily sample the same type of galaxies since the different filters 
sample different rest-frame wavelength ranges. 
While the results of van~Dokkum et al.\ prove that high-redshift red 
galaxies can be bright, their spectroscopic sample (six of eleven 
$J-K_{s}$-selected galaxies) may be biased toward systems with ongoing 
star formation.
 
The average ERO fraction in our quasar fields is above 20 \%, and 
for the four richest fields in our sample the fraction is as high
as $\approx50$ \%. This is what we expect if the fields contain clusters or 
concentrations of galaxies with an already existing relation between
morphology and density \citep{dressler97}. 
This is also the suggestion
from the work by \citet{best03} who find that red galaxies are concentrated
toward $z\approx1.5$ powerful radio sources.

Whereas the magnitudes and colors of the EROs are consistent with models
of passively evolving galaxies, another possibility is that they might be 
starbursting systems reddened by dust. Hence, an alternative explanation
is that the quasars live in dense environments dominated by galaxies with 
enhanced star formation. In this case, the excess EROs may be dusty starburst 
galaxies, possibly the progenitors of massive ellipticals in a forming 
cluster. There are examples of fields around high-redshift radio-loud AGN 
with excesses of starforming galaxies such as Ly-alpha emitters 
\citep{kurk00,pentericci00,venemans02}. 
Also, \citet{ivison00} find a concentration of submm sources in the field of 
the $z=3.8$ 
radio galaxy 4C 41.17 raising the possibility that the EROs in this field are 
dusty starburst systems rather than passively evolving galaxies. 
This possibility is also discussed by \citet{smail03}
for a luminous submm source in the field of the radio galaxy 53W002 at $z=2.4$.

One might expect a correlation between the radio properties of the quasars
and their ERO excesses if there is e.g.\ a relation between radio power
and black hole mass \citep{wold00,laor00,lacy01}. We have checked this for
our sample, but do not find that the radio power correlates with the ERO
excess. 

\subsection{EROs in foreground lensing structures}

Little is known yet of the luminosity function of EROs. In the spectroscopic
survey by \citet{cimatti02}, which is 67 \% complete at $K_{s}\leq19.2$ mag, 
the $R-K_{s} \geq 5.0$ mag EROs are found to lie at $z=0.7-1.5$. The redshift 
distribution derived by \citet{miyazaki02} based on photometric redshifts is 
in broad agreement with Cimatti et al's, but has a tail toward higher 
redshifts. If we assume that our ERO sample has a redshift distribution 
similar to that 
found by Cimatti et al.\ and Miyazaki et al., we have to explain why the ERO 
surface density in the quasar fields is so significantly different from that 
in the large random-field surveys. We are sampling the ERO distribution with
several small, but widely separated fields, 
and should therefore be minimally affected by cosmic variance. 
Instead, the opposite seems to be the case; the quasar fields 
sample the densest regions in the ERO distribution. A large number 
of the quasar fields contain clumps of EROs, similar to what is seen only in 
the densest areas in the large random-field surveys. 
If the EROs have a peak redshift of $z\approx1.0$, they must therefore be
clusters of galaxies, or overdensities in the ERO distribution, lying
along the lines of sight to the quasars. 

A biasing of quasar lines of sight with respect to $z\approx1$ 
overdensities in the ERO distribution can be explained by 
gravitational lensing. 
If we assume that the overdensities in the quasar fields do not extend much 
beyond the edges of the fields their typical scale is $\sim 1$ arcmin, 
corresponding to 2--3 Einstein radii for galaxy clusters at $z\sim1$.
The most efficient lens redshift for a quasar at $z=2$ is 0.5--0.6, but
a magnification can still occur if the lens lies at $z\sim0.8-1.0$.

The quasars in our sample are very luminous, eight of them 
have $M_{V}<-27.2$ mag ($H_{0}=50$ km\,s$^{-1}$\,Mpc$^{-1}$), 
and hence classify as highly luminous quasars.
They therefore fulfill the three criteria for maximizing the probability of 
a magnification bias, i.e.\ they are 
distant, bright and have a 
steep number-magnitude slope (Claeskens \& Surdej 2002, and references
therein). If a magnification bias is the explanation for our results, we 
might expect that the fields around the most luminous quasars contain the 
highest overdensities of EROs, but our data do not show any significant 
correlations between quasar luminosity and ERO surface density to support
this.

Furthermore, if the majority of the EROs are associated with 
overdensities along the line of sight to the quasars, we might expect
the spectra of the quasars in the richest fields to have strong (rest-frame
equivalent width $\gtrsim 0.7$ {\AA}) metal 
absorption line systems \citep{liu00}. We have found absorption line data
for seven of the 13 quasars in our sample \citep{junkkarinen91,junkkarinen92}.
Two of the four richest fields have such data; Q2150$+$053 show no strong 
metal absorption lines and Q2338$+$042 have 
CIV and FeII at $z_{abs}\approx1.80$ \citep{barthel90}.
Other quasars with metal absorption line systems in the foreground
is Q0830$+$115 with FeII at $z_{abs}=0.92$ and MgII 
at $z_{abs}=0.80$ \citep{sargent89}, Q0941$+$261 with MgII
systems at $z_{abs}=0.71$ and $z_{abs}=1.09$, and Q1354$+$258 with 
MgII at $z_{abs}=0.86$ and some weaker FeII and MgII
systems at $z_{abs}=1.42$ \citep{barthel90}, but these quasars lie 
in fields with relatively few EROs compared to the richer fields. 
The data for Q0758$+$120 and Q1629$+$680 show no strong metal absorption
line systems \citep{barthel90}. There is thus no clear trend of increasing
number of absorption line systems with ERO richness. Also, the quasars
in our sample  
were not selected on the basis of having strong or numerous 
absorption line systems. 

\section{Conclusions}
\label{section:sec6}

On the average, the quasar fields studied here have a cumulative 
surface density of EROs at $K_{s}\leq18$, 18.5 and 19 mag which is
two-three times larger than what is found in the general field. The excess is 
significant at the $\approx3\sigma$ level even after taking into account that 
the ERO distribution is strongly clustered. There is also an excess of EROs at
fainter magnitudes, but at $K_{s}\approx20$ mag our counts become strongly 
affected by incompleteness and a secure quantitative estimate of the excess for
$K_{s}>20$ mag is difficult. However, by making a correction for the 
incompleteness, we find that the ERO excess at $K_{s} \leq 19.5$ and 
$K_{s}\leq20$ mag may also be significant at the $\ga 3\sigma$ level.

The quasar fields are small, typically 2.1$\times$2.1 arcmin$^{2}$
(limited by the near-infrared field of view), but
widely separated, and therefore sample the ERO population with minimal
influence from cosmic variance. The fact that the excess is 
statistically significant
therefore suggests that the quasar fields probe the densest 
regions in the ERO distribution, implying that the fields around luminous, 
high-redshift quasars are on the average, overdense in EROs.
There are two possible explanations for the observed overdensities.
(1) The majority of the EROs may be associated with concentrations
of galaxies at the quasar redshifts, or (2) the EROs lie in the foreground,
along the line of sight to the quasars. 

Based on both spectroscopic observations of $z>2$ red galaxies
\citep{vandokkum03} and spectral synthesis models \citep{frv97}, we 
argue that a redshift of $z\approx2$ for the EROs is plausible. 
The $K_{s}\approx19$ mag ERO excess may therefore be physically 
associated with the quasars, in which case their magnitudes and colors are 
consistent with $\gtrsim L^{*}$ passively evolving galaxies. 

The EROs at $K_{s}\geq19-19.5$ mag are more clustered in the immediate
20 arcsec region around the quasars, hence the radial distribution 
supports the explanation that the fainter EROs are physically associated with 
the quasars. Furthermore, the fraction of EROs among the $K_{s}$-selected 
galaxies
in the quasar fields is unusually high, typically 20--25 \%, in contrast
to $\lesssim 10$ \% which is found in random-field surveys 
\citep{daddi00,ss00,roche02}. This is consistent
with the ERO excess being associated with dense concentrations of 
red, possibly evolved, galaxies. The high ERO fractions 
may therefore indicate that a morphology-density relation 
\citep{dressler97} exists in the quasar fields.
The candidate ERO clusters in the quasar fields may therefore
represent the oldest and most evolved systems at $z\gtrsim 2$, and will
be crucial for testing models of structure and galaxy formation. 
If the ERO excess is dominated by dusty starforming galaxies instead of
passively evolving galaxies, the EROs might be the progenitors of massive 
ellipticals in a forming cluster associated with the quasars. 

The second explanation, that the ERO excess lies in the foreground, is 
equally interesting because it implies that the lines of sight toward 
high-redshift, luminous quasars are biased toward overdensities in the ERO 
distribution. In this case, the
ERO excess is most likely associated with clusters of galaxies at
$z\approx1$ giving rise to a magnification bias of the quasars. In this
interpretation, we would be seeing candidate clusters 
with passively evolving galaxies at $z\approx1$, a redshift where 
still only less than a handful of clusters are known. However, we see no 
trend of quasar brightness with ERO overdensity as might be expected in
this interpretation. Also, there is no association of the numbers
of absorption line systems with ERO richness for seven of the 13 quasars 
with absorption line data. 

With the current data, it is not possible to tell whether the ERO excess
is part of a small-scale overdensity centered on, or lying 
close to, the quasars, 
or if it is part of a larger structure that extends over a wider area. 
To address this issue, we have begun an imaging program using $J$- and 
$K_{s}$-filters at the Palomar 200-inch telescope over much larger fields.
With these data we will be able to trace the ERO population around
the $z\approx2$ quasars on scales up to 4--5 Mpc. 

\acknowledgments

We are grateful to Dave Thompson, Lin Yan and Mark Lacy for helpful
discussions, and to the referee for a careful review of the manuscript. 
This publication makes use of data products from the Two Micron All Sky
Survey, which is a joint project of the University of Massachusetts and
the Infrared Processing and Analysis Center/California Institute of 
Technology, funded by the National Aeronautics and Space Administration
and the National Science Foundation. 
This research has also made use of the NASA/IPAC Extragalactic Database 
(NED) which is operated by the Jet Propulsion Laboratory, California 
Institute of Technology, under contract with the National Aeronautics and 
Space Administration.

\end{document}